\documentstyle[aps,manuscript]{revtex}

\begin{document}

\title{Evolutionary minority game with heterogeneous strategy distribution}
\author{T.S. Lo$^{1}$, S.W. Lim$^{1}$, P.M. Hui$^{1}$, and N.F.
Johnson$^{2}$}
\address{$^{1}$ Department of Physics, The Chinese University of Hong
Kong,\\
Shatin, New Territories, Hong Kong.\\
$^{2}$ Department of Physics, University of Oxford, \\ Clarendon Laboratory,
Oxford OX1 3PU, England, UK.}

\maketitle
\begin{abstract}
We present detailed numerical results for a
modified form of the so-called Minority Game, which provides a simplified model
of a competitive market. Each agent has a limited set of strategies, and
competes to be in a minority. An evolutionary rule for strategy modification is
included to mimic simple learning. The results can be understood by considering
crowd formation within the population.

\end{abstract}


\newpage
\section{introduction}

There is rapidly growing interest
in the study of complex adaptive systems (CAS)\cite{holland}.
They not only provide a challenging problem for physicists because of the
non-trivial self-organizing phenomena which can emerge, but
also have potential applications in a variety of
economical, biological and financial
problems\cite{arthur,stanley,confs}.
An important step forward in agent-based models of CAS
was made by Challet and
Zhang\cite{challet,savit} who proposed the
so-called Minority Game (MG) in which an odd number of agents
successively compete to be in the minority.
The agents make decisions by
evaluating the performance of their
strategies from past experience and hence they can adapt.
The strategies are randomly assigned to the agents in the beginning
of the game and are used throughout the game, hence introducing
some quenched disorder.
The agents have access to
global information, which is in turn generated by the
actions of the agents themselves.  As the game progresses,
non-trivial fluctuations arise in the agents' collective decisions -
these can be understood in terms of the dynamical formation of crowds
consisting of agents using correlated strategies, and anticrowds
consisting of agents using the anticorrelated strategies\cite{crowd}.
Challet and coworkers have recently presented a
remarkable connection between the MG and spin
glass systems\cite{spinglass}.

The MG, however, does not incorporate evolution.
Agents may get stuck with poorly performing
strategies as a result of the initial (random) strategy distribution.
Johnson {\em et al}\cite{prl,royal,HLJ,LHJ}
proposed a version of MG which involves an evolving
population.  In this so-called evolutionary
minority game (EMG), all the agents hold one and
the same strategy which is simply
to follow the most recent trend.  Hence the strategy is dynamical.
Each agent also carries a probability $p$ characterizing the
chance of following the prediction of the strategy.  Evolution
comes in by allowing
agents to modify their $p$ values when their success
rate becomes too low, hence mimicking the notion that market participants ought
to learn from past mistakes.  Surprisingly, agents who either always follow or
never follow the trend generally perform better than cautious agents
\cite{prl}.

In an effort to explain the non-trivial behaviour observed in MG and
EMG, D'hulst and Rodgers\cite{dHR} pointed out the relevance of the
Hamming distance between strategies and studied a modified version
of EMG in which each agent holds a randomly selected strategy and
a probability $p$. The theory gives results which appear to resemble those
of EMG.  In contrast to EMG, however, each agent now has a fixed (i.e.
non-dynamical) strategy throughout the game.   Moreover, it has recently
been
pointed out that this modified model and  the basic EMG actually give
qualitatively different numerical results\cite{HLJ}.
Recently, we presented a theory
for the basic EMG which gives good agreement with
numerical data\cite{LHJ}. Our theory properly includes the
self-interaction of the agents \cite{LHJ}.

The modified EMG model of D'hulst and Rodgers\cite{dHR} is, however, an
interesting model in its own right. In particular, it brings together the
idea of
evolution from the basic EMG and the idea of random initial strategy
distribution from the MG.  Here we present detailed
numerical results for this modified EMG model of D'hulst and
Rodgers\cite{dHR}.
The results are contrasted with those for the basic EMG and MG where
possible.
The paper is organized
as follows.  The modified EMG model is defined in Sec.II.
Section III gives the numerical results.  Differences and similarities are
pointed out between the modified model and the basic EMG and MG.
Our results are summarized in Sec.IV.

\section{model}

The model consists of an odd number $N$ of agents.
Each agent has to choose
between two decisions, 0 or 1, at each timestep.
The winning side, i.e., the minority side, represents
the outcome of the game at that timestep.  These outcomes form the global
information  made known to all agents.  As in the MG, the agents make their
decision based on the most recent $m$ outcomes.
There are a total of $2^m$ different possible histories.
A strategy is defined as a mapping from the history space
to the action space.  Since for each possible history there are
two possible decisions, there exist a total of
$2^{2^m}$ different strategies.
Each agent is allowed to pick {\em one}
strategy from the pool of strategies in the beginning of the game.
Some agents may share a common strategy.
The strategy remains fixed throughout the game.  This
random assignment of strategies is identical to that in MG\cite{challet},
but different from the EMG\cite{prl}.  As in the EMG, however,
evolutionary behaviour is allowed through
the assignment of a
parameter $p$ to each agent characterizing
the probability that the agent follows the prediction of his strategy.
The agent thus has a probability $1-p$ to make the decision opposite
to what his strategy predicts.
Each agent is randomly assigned a $p$-value in the beginning
of the game.
 The scores of all agents are set to zero initially.
An agent wins and gains one point if he belongs to the minority group.
Otherwise, he loses with one point deducted.
If the score drops to $d$ ($d < 0$),
an agent is allowed to modify his $p$ value by choosing a
value within a range $R$ centered at the
original $p$ value and his score is reset to zero.
Reflective boundary conditions ensure that
$p$ always lies within the range $0 \leq p \leq 1$.
Results are found to be insensitive to the particular
choice of boundary conditions.

\section{results}

Figure 1 shows the distribution of the $p$-values $P(p)$ among
the agents, obtained numerically after the transient stage of the game has
died
away.   The distribution is normalized such that $\int_{0}^{1} P(p) dp = N$.
The parameters chosen are $N=101$, $d=-4$ and $R=0.2$.  The features
observed are insensitive to the initial $p$-distribution.  It is
observed that $P(p)$ depends sensitively on $m$, in contrast to
the $m$-independence of $P(p)$ observed in the basic EMG\cite{prl,HLJ,ceva1}.
For small $m$,
$P(p)$ has peaks near $p\sim 1$ and $p\sim 0$ implying agents who always
act according to or opposite to the trend perform better than those with
intermediate values of $p$.  This feature is qualitatively
similar to that in the basic EMG with the $m=1$ results closely
resembling those reported in Ref. \cite{prl}.  For small $m$ such that
$2\cdot 2^m \ll N$, the strategies are almost uniformly distributed among
the agents and all strategies are played.  This distribution of strategies,
together with  the fact that every agent holds only one strategy, therefore
ensures that every  strategy and its anticorrelated partner will be used in
each
turn of  the game.  The form of $P(p)$ for small $m$, namely symmetrical
about
$p=0.5$ with peaks at the extreme values on either side, leads to
good cancellation between the actions taken by the
agents using anti-correlated pairs of strategies\cite{crowd}.
This self-organization
in the population, in turn, has the advantage that it increases
the number of winning agents per turn.  Although this optimization
of a `global profit' is not the aim of each individual agent when
the decision is made, it results from the competition among the
agents.  As $m$ increases, $P(p)$ gradually flattens off.  In the
limit of large $m$, i.e. $2\cdot 2^m \gg N$, only a small portion
of the whole pool of strategies is picked by the agents.  It is
therefore unlikely that strategies which are anti-correlated to each
other are being played.  With or without the effect of the
$p$-values, the game is
in the random coin-toss limit.  There is no advantage to having one
particular $p$-value over another, hence the flat form of $P(p)$ at high
$m$.
The inset in Fig. 1 gives the
$m$-dependence of the mean lifespan $L(p)$, which is the average
number of turns a certain value $p$ survives between modifications.  The
features are similar to those in $P(p)$.

The above discussion implies that for small $m$, the standard
deviation (SD) in the number of agents making a certain decision (either
0 or 1) is small due to the cancellation in the actions taken by
agents using anti-correlated pairs of strategies.  For large $m$, the
SD approaches the random coin-toss limit of $\sqrt{N}/2$.  Figure 2
shows the SD as a function of $m$ for $N=101$.  It can be seen that the SD
does indeed increase  monotonically with $m$, up to the random coin-toss
limit.
Also included in Fig. 2 are  the results, averaged over different initial
distributions of strategies, for the minority game (MG) with
$s=2$.
These $s=2$
MG results are included in order to contrast the different ways
in which adaptive behaviour is introduced in the two models.  In the $s=2$
MG, each agent randomly picks two strategies initially, with repetitions
allowed.  The picked strategies may not always give the opposite action
in response to a given history of the most recent $m$ outcomes.  In the
modified EMG, each agent effectively holds one randomly picked
strategy together with its anti-correlated counterpart, 
and chooses between them
stochastically using $p$.  
The modified EMG is hence effective in forming
similar-sized crowds and anticrowds, thereby yielding a smaller-than-random
SD for a wide range of $m$ values (see Fig. 2).   The
$s=2$ MG, however,
has a much  higher SD for small $m$ as it does not have the built-in
crowd-anticrowd cancellation effect. Also shown in Fig. 2 is the SD for the 
basic EMG: the SD is $m$-independent and takes on
a value  close to the $m=1$ result of the modified EMG.  
In the basic EMG,
every  agent carries the same strategy at a given moment and the
self-organized
distribution $P(p)$ leads to an effective crowd-anticrowd cancellation which
is
independent of the size of the strategy space, and hence $m$.

Figure 3 shows the dependence of $P(p)$ and $L(p)$ (inset) on $d$.
The properly normalized $P(p)$ as shown does not depend on $d$ while
$L(p)$ depends on $|d|$ in such a way that $L(p) \sim |d|$.  Both
features are identical to those in the basic EMG.
If we denote $\tau(p)$
as the average winning probability of an agent playing with value $p$,
$L(p) = |d|/(1-2\tau(p))$.  It should be noted that for all versions
of the MG, $\tau < 1/2$.  However, the results of $L(p)$ reported here give
values of $\tau(p)$ which are quite different from those obtained by
numerically solving the set of equations given in
Ref. \cite{dHR}.   
Figure 4 shows  $P(p)$ and $L(p)$ for various values of $R$.  Both $P(p)$ 
and $L(p)$ show some dependence on $R$ with the peaks on both sides becoming
less pronounced as $R$ increases.  Comparing with the results of EMG\cite{HLJ}, 
the modified EMG has a more sensitive dependence on $R$ although 
the results are qualitatively similar.  A full explanation 
of the enhanced sensitivity to $R$ requires us to consider the 
details of the dynamics of the game, since $R$ controls 
the effective diffusion in $p$-space.  This will be addressed 
in future work.

\section{Discussion}

We have presented numerical results for a
modified EMG.  The distribution $P(p)$ is found to be $m$-dependent,
in contrast to the basic EMG.  For $m=1$,
the modified EMG gives results similar to the basic EMG.
As $m$ increases, the strategy distribution leads to a larger SD
than the basic EMG since the crowd-anticrowd cancellation effect
gets reduced.  For large $m$, the strategy pool is so large that
the game is effectively in the random coin-toss
limit.  In terms of the behavior of the SD, the
present modified EMG interpolates between the basic EMG at small $m$, and
the MG at large $m$.  We also found that the $R$ dependences 
of $P(p)$ and $L(p)$ are enhanced by the implementation of the random 
strategy distribution. 

In both the basic EMG and the modified EMG, the adaptability
associated with the stochastic
$p$ parameter reduces  the SD to values {\em below} the random coin-toss
limit for
all values of
$m$.  This was reported earlier \cite{prl} for the basic EMG:
interestingly, another example of reduction of SD to values below the random
coin-toss limit was subsequently reported for the so-called Thermal Minority
Game (TMG)
\cite{TMG1,garrahan,TMG2}. In the TMG, each agent may use one of his $s$ strategies
in
each turn according  to a probability determined by a parameter $T$ playing
the
role of  `temperature'\cite{TMG1,TMG2}.   It was found that for $s=2$, for
example, the SD for the TMG at small $m$  drops to values below the
coin-toss
limit.  The reason can be understood quantitatively in terms of
crowd-anticrowd
formation
\cite{theta}. In essence, the built-in frustration in the basic MG due to
the
distribution of strategies among agents gives rise to a large SD at small
$m$
because of the large size of the crowds as compared to the
anticrowds\cite{crowd}.  The introduction of temperature  effects helps to
build
up the crowd-anticrowd cancellation by allowing similar-sized crowds and
anticrowds to form\cite{theta},  hence reducing the SD.  The present
modified EMG
is similar to the TMG in that
 each agent can be regarded as holding  effectively two strategies, together
with a stochastic rule for strategy-use at each timestep: one strategy is
randomly  picked and played with probability
$p$ while the anti-correlated strategy is played with probability $1-p$.  Thus,
$p$ plays a somewhat similar role  to the temperature $T$ in TMG. (Compare Fig.
2 of the
present paper with the inset of Fig. 3 in Ref. \cite{garrahan}).
However, we emphasize that the ($s=2$) strategies effectively  held
by agents in the modified EMG  {\em always} form an anti-correlated pair -
this is
not the case in TMG. 
Finally, we note that a variant of the MG was proposed by 
Challet {\em et al} in which agents also have $s=2$ anti-correlated
strategies\cite{anti}: however  the agents did not 
use a stochastic variable to choose between their strategies at each timestep,
hence the results are not directly related to those of the present modified EMG
model.

\begin{figure}

\caption{The frequency distribution $P(p)$ and average lifespan
$L(p)$ (inset) of the modified EMG
for $N=101$, $d=-4$, $R=0.2$ and different values
of $m$ ($m=1$, $4$, $7$, $10$). In contrast to basic EMG, both
$P(p)$ and $L(p)$ depend on $m$.}
\vspace*{0.3 true in}

\caption{The standard deviation (SD) in the number of agents
making a certain decision as a function of $m$ for the
modified EMG, the basic EMG, and the MG with $s=2$. $N=101$, $d=-4$ and
$R=0.2$.}
\vspace*{0.3 true in}

\caption{The frequency distribution $P(p)$ and average lifespan
$L(p)$ (inset)
for $N=101$, $R=0.2$, $m=3$ and different values
of $d$ ($d=-1$, $-4$, $-7$, $-10$).}
\vspace*{0.3 true in}

\caption{The frequency distribution $P(p)$ and average lifespan
$L(p)$ (inset)
for $N=101$, $m=3$, $d=-4$, and different values
of $R$ ($R=0.5$, $1.0$, $1.5$, $2.0$).}
\vspace*{0.3 true in}

\end{figure}


\newpage
\begin{thebibliography}{99}

\bibitem{holland} J.H. Holland, {\em Emergence: From chaos to order},
(1998) (Addison-Wesley, Reading).

\bibitem{arthur} W.B. Arthur, Amer. Econ. Rev. {\bf 84}, 406 (1994);
Science {\bf 284}, 107 (1999).

\bibitem{stanley} H.E. Stanley, Computing in Science \& Engineering
Jan/Feb, 76 (1999); Physica A {\bf 269}, 156 (1999).

\bibitem{confs} See the proceedings of the International Workshop  on
Econophysics and Statistical Finance in Physica A {\bf 269}, 1-183
(1999).


\bibitem{challet} D. Challet and Y.C. Zhang, Physica A {\bf 246}, 407
(1997);
{\em ibid.} {\bf 256}, 514 (1998); {\em ibid.} {\bf 269}, 30 (1999).


\bibitem{savit} R. Savit, R. Manuca and R. Riolo, Phys. Rev. Lett.  {\bf
82}, 2203
(1999).

\bibitem{crowd} N.F. Johnson, M. Hart and P.M. Hui, Physica A {\bf 269}, 1
(1999); M. Hart, P. Jefferies, N.F. Johnson and P.M. Hui,
cond-mat/0003486.

\bibitem{spinglass} 
D. Challet and M. Marsili, Phys. Rev. E {\bf 60}, R6271
(1999); D. Challet, M. Marsili, and R. Zecchina,
Phys. Rev. Lett. {\bf 84}, 1824 (2000);
D. Challet and M Marsili, cond-mat/9908480.

\bibitem{prl} N.F. Johnson, P.M. Hui, R. Jonson and T.S. Lo,  Phys. Rev.
Lett. {\bf 82}, 3360 (1999).

\bibitem{royal} N.F. Johnson, P.M. Hui and T.S. Lo, Phil. Trans.  Royal Soc.
London A {\bf 357}, 2013 (1999).

\bibitem{HLJ} P.M. Hui, T.S. Lo, and N.F. Johnson, cond-mat/0003309.

\bibitem{LHJ} T.S. Lo, P.M. Hui, and N.F. Johnson, cond-mat/0003379.

\bibitem{dHR} R. D'hulst and G.J. Rodgers, Physica A {\bf 270}, 514 (1999).

\bibitem{ceva1} E. Burgos and H. Ceva, cond-mat/0003179.

\bibitem{TMG1} A. Cavagna, J.P. Garrahan, I. Giardina and D. Sherrington,
Phys. Rev. Lett. {\bf 83}, 4429 (1999). 

\bibitem{garrahan} J.P. Garrahan, E. Moro and
D. Sherrington, cond-mat/0004277.

\bibitem{TMG2} D. Challet, M. Marsilli and R. Zecchina, cond-mat/0004308.

\bibitem{theta} M. Hart, P. Jefferies, N.F. Johnson and P.M. Hui,
cond-mat/0004063; P. Jefferies, M. Hart, N.F. Johnson and P.M. Hui,
cond-mat/0005043.

\bibitem{anti} See Sec. IV of D. Challet, M. Marsili and Y.C. Zhang,
cond-mat/9909265. 

\end{thebibliography}
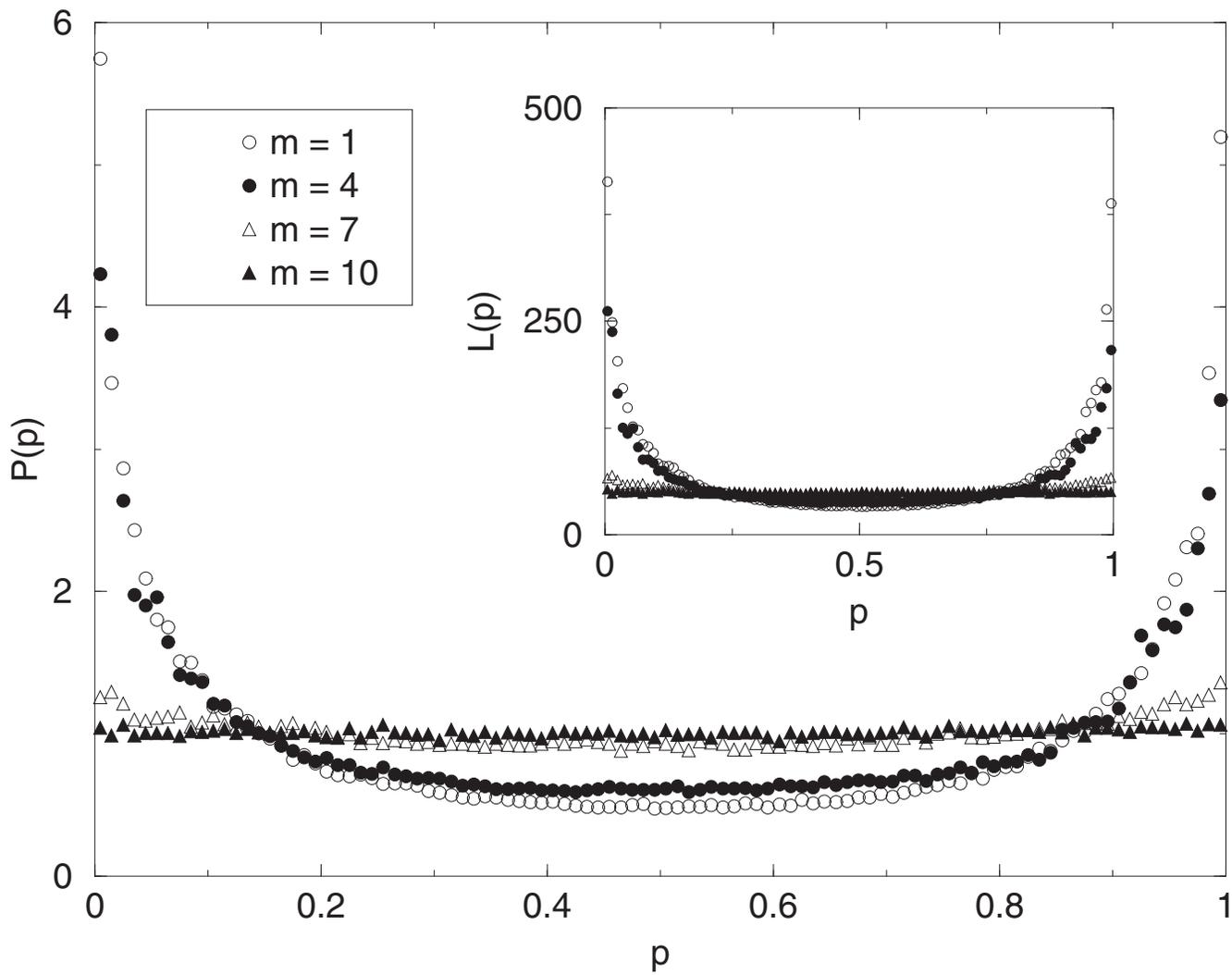
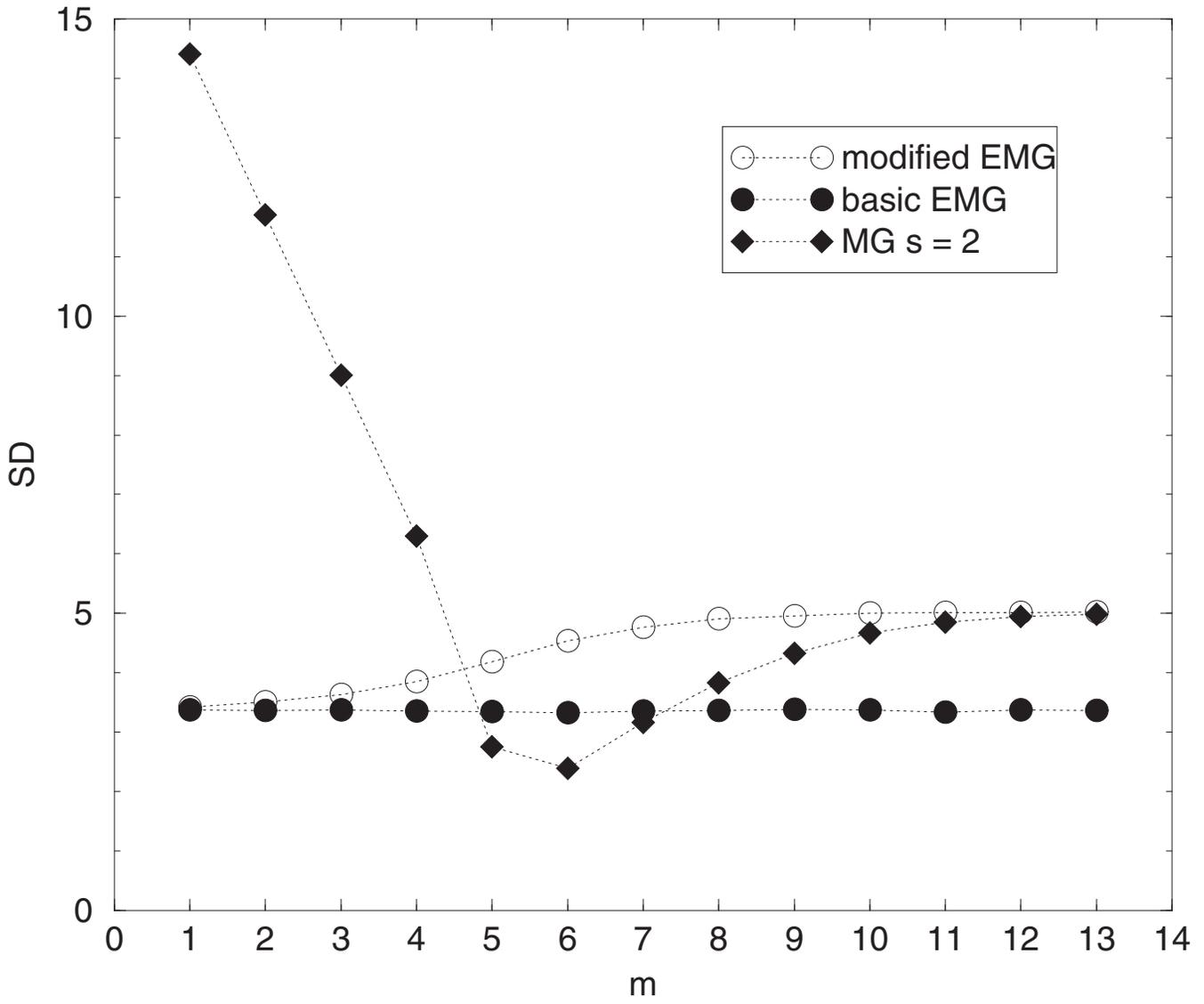
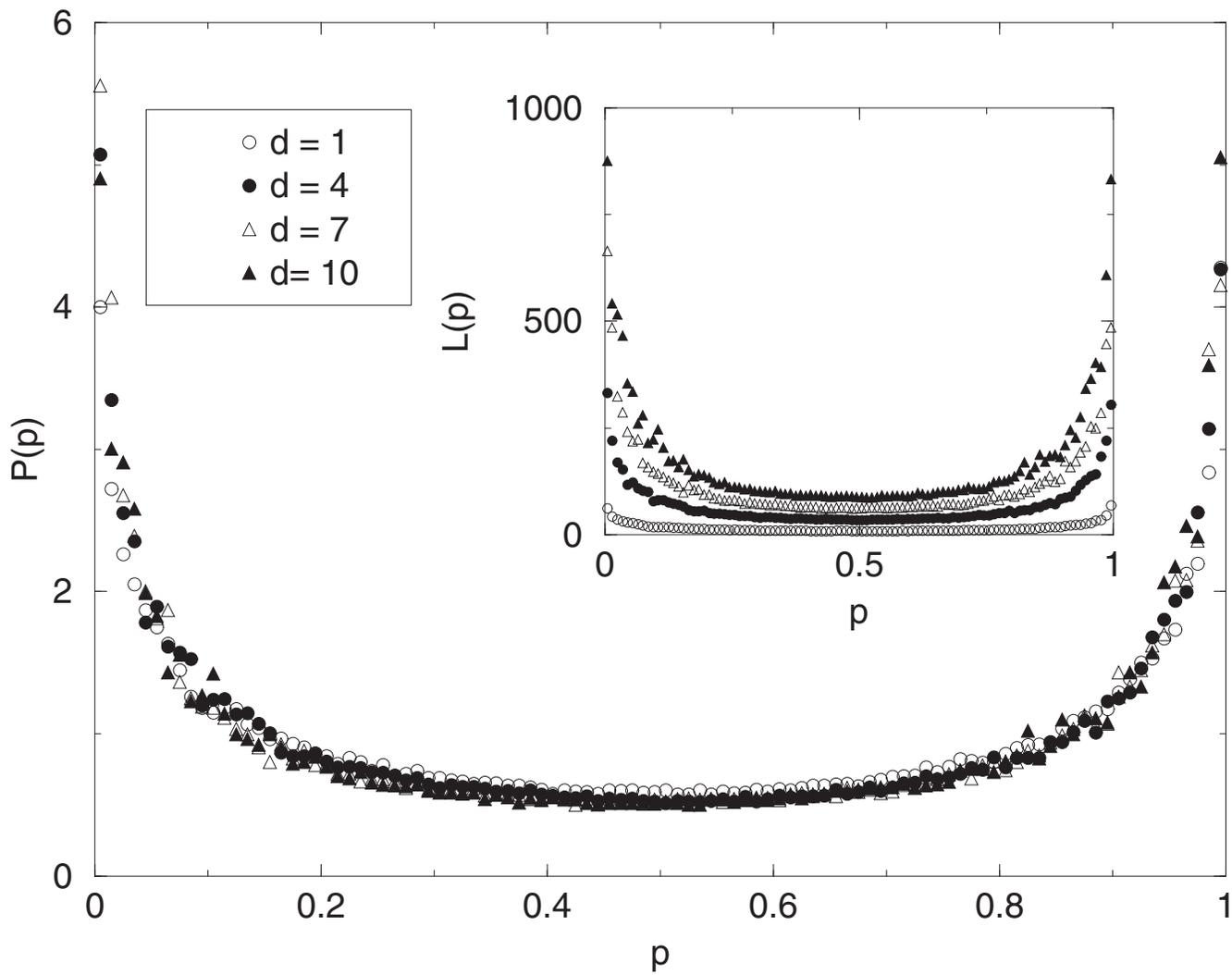
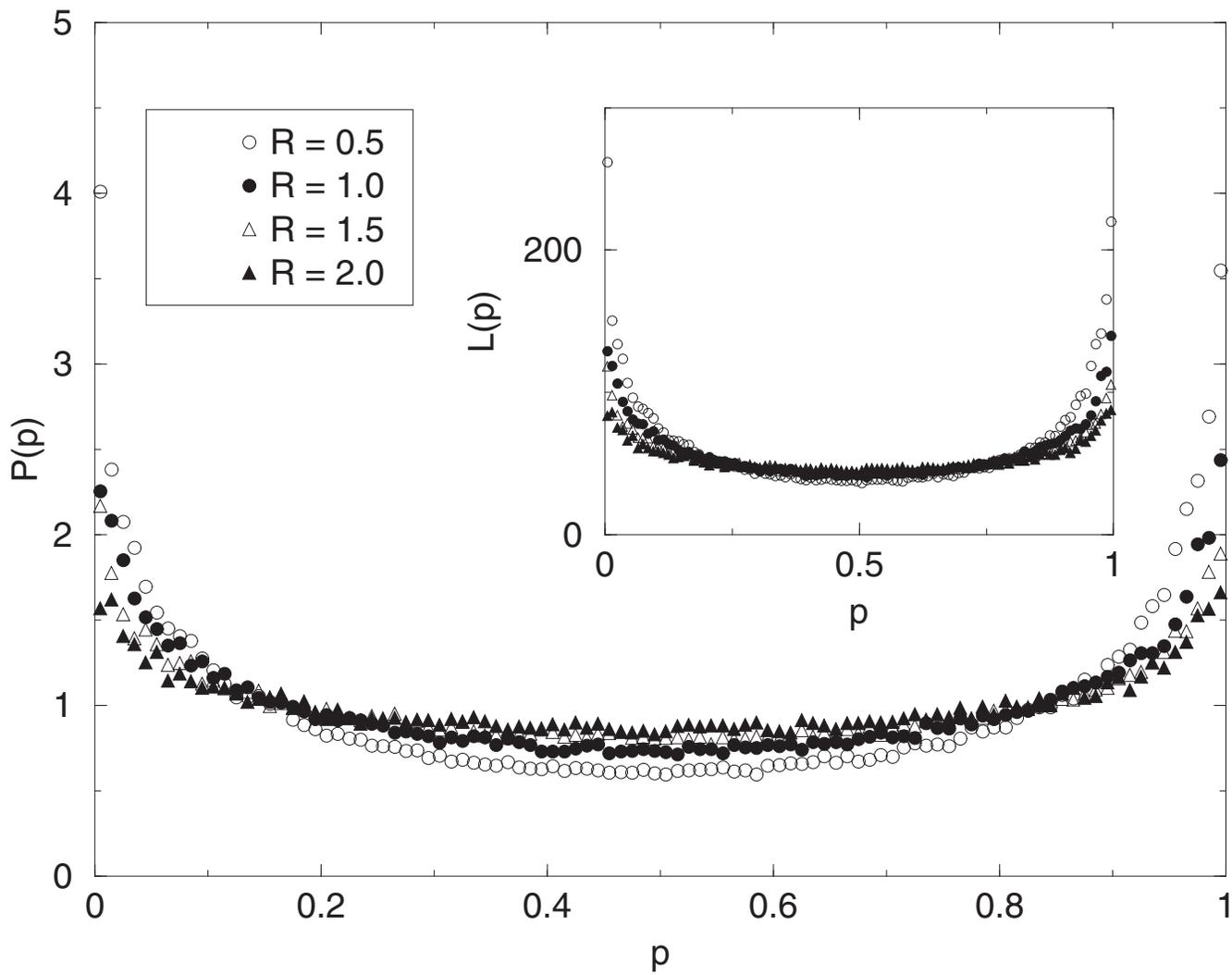
\end{document}